\documentclass[12pt]{article}
\usepackage{graphics}
\usepackage{amssymb}
\usepackage{amsmath}

\usepackage{bm}

\newcommand{\rf}[1]{(\ref{#1})}
\newcommand{\beq}{\begin{equation}}
\newcommand{\beql}[1]{\beq\label{#1}}
\newcommand{\eeq}{\end{equation}}
\newcommand{\bea}{\begin{eqnarray}}
\newcommand{\eea}{\end{eqnarray}}

\renewcommand{\d}{\mbox{d}}
\newcommand{\g}{\gamma}

\newcommand{\lam}{\lambda}
\newcommand{\La}{\Lambda}
\renewcommand{\b}{\beta}
\renewcommand{\a}{\alpha}

\newcommand{\Del}{\Delta}

\renewcommand{\k}{\kappa}

\newcommand{\ra}{\rangle}
\newcommand{\la}{\langle}
\newcommand{\prt}{\partial}

\newcommand{\tV}{{\tilde{V}}}

\begin{document}

\begin{center}
\vspace{24pt}
{ \Large \bf CDT meets Ho\v rava-Lifshitz gravity}

\vspace{30pt}

{\sl J. Ambj\o rn}$\,^{a,c}$,
{\sl A. G\"{o}rlich}$\,^{b}$,
{\sl S. Jordan}$\,^{c}$,
{\sl J. Jurkiewicz}$\,^{b}$
and {\sl R. Loll}$\,^{c}$

\vspace{24pt}
{\footnotesize

$^a$~The Niels Bohr Institute, Copenhagen University\\
Blegdamsvej 17, DK-2100 Copenhagen \O , Denmark.\\
{ email: ambjorn@nbi.dk}\\

\vspace{10pt}

$^b$~Institute of Physics, Jagellonian University,\\
Reymonta 4, PL 30-059 Krakow, Poland.\\
{ email: atg@th.if.uj.edu.pl, jurkiewicz@th.if.uj.edu.pl}\\

\vspace{10pt}

$^c$~Institute for Theoretical Physics, Utrecht University, \\
Leuvenlaan 4, NL-3584 CE Utrecht, The Netherlands.\\
{ email: s.jordan@uu.nl, r.loll@uu.nl}\\

\vspace{10pt}
}
\vspace{48pt}

\end{center}


\begin{center}
{\bf Abstract}
\end{center}
The theory of causal dynamical triangulations (CDT) attempts to define
a nonperturbative theory of quantum gravity 
as a sum over space-time geometries.
One of the ingredients of the CDT framework is  
a global time foliation, which also plays a central role in 
the quantum gravity theory recently formulated by Ho\v rava.
We show that the phase diagram of CDT bears a striking resemblance with 
the generic Lifshitz phase diagram appealed to by Ho\v rava.
We argue that CDT might provide a unifying nonperturbative framework
for anisotropic as well as isotropic theories of quantum gravity.

\newpage

\section{Introduction}\label{intro}

A major unsolved problem in theoretical physics is to reconcile
the classical theory of general relativity with quantum mechanics.
Recently there has been a resurgence in interest in using mundane quantum field 
theory to address this question. Progress over the last ten years
in the use of renormalization
group (RG) techniques \cite{RG} 
suggests that the so-called asymptotic safety scenario,
originally put forward by S. Weinberg \cite{weinberg}, 
may be realized, namely, the existence of 
a nontrivial ultraviolet fixed point, where one can define
a theory of quantum gravity. 

In tandem with this approach, the method
of Causal Dynamical Triangulation (CDT) has been developed, likewise
with the aim of defining and constructing a nonperturbative quantum gravity
theory \cite{emerge,blp,semi,agjl,bigs4} (for recent reviews,
see \cite{recent}). CDT provides a lattice framework in which a variety of 
nonperturbative field-theoretical aspects of quantum gravity can be studied,
including in principle predictions from other candidate theories. 
Despite the fact that the CDT and the RG approaches use rather different
sets of tools, they might be two sides of the same coin. Locating a suitable
UV fixed point in causal dynamical triangulations would provide strong
evidence that this is indeed the case and that ``asymptotic safety" is on the right track. 

More recently, P.\ Ho\v rava has
suggested yet another field-theoretical approach to quantum 
gravity in the continuum \cite{horava}, since dubbed Ho\v rava-Lifshitz gravity,
where the four-dimensional diffeomorphism symmetry of general relativity is explicitly
broken. Assuming a global time-foliation, time and space are treated differently, in
the sense that only suitable {\it second}-order derivatives in time appear to render
the quantum theory unitary, while higher-order spatial derivatives ensure
renormalizability. 

A common key ingredient in both CDT and Ho\v rava-Lifshitz gravity is a global 
time foliation, with the difference that in CDT this is not directly associated with
a violation of diffeomorphism symmetry, since the dynamics is defined directly
on the quotient space of metrics modulo diffeomorphisms. This raises the
question whether new insights can be gained by analyzing and interpreting
CDT quantum gravity
in a generalized, anisotropic framework along the lines of Ho\v rava-Lifshitz gravity.
The reference frame until now has been a covariant one, assuming that any
UV fixed point found in the CDT formulation could be identified with that found
in the covariant renormalization group approach, appealing to the general sparseness 
of fixed points\footnote{Of course, one  should also show that a lattice fixed point 
reproduces the critical exponents of the RG treatment.}.  
At the same time, we have presented general arguments in favour of
a reflection-positive transfer matrix 
in the (Euclideanized version of) CDT \cite{d4prl,d4}.
Thus the conditions for a unitary quantum field theory at the UV fixed
point are also met. The philosophy behind
formulating gravity at a Lifshitz point was that unitarity in a theory
of quantum gravity should be the prime requirement, rather than treating 
space and time on the (almost) equal footing required by special relativity.
We conclude that the CDT approach not only shares the time-foliated structure
of spacetime, but also the enforcement of unitarity by construction with 
Ho\v rava-Lifshitz gravity. 

This led us to asking whether CDT may be able to capture aspects of the latter,
despite the fact that no higher-order spatial derivative terms 
are put in by hand in the CDT action. Some support for this idea comes from
the fact that one UV result which can be compared explicitly, namely, 
the nontrivial value of
the spectral dimension of quantum spacetime, appears to coincide in both 
approaches \cite{spectral,spectral-hl}. 
Interestingly, also the renormalization group approach was able to reproduce the
same finding, after the spectral dimension had first been measured in simulations of
CDT quantum gravity, a result taken at the time as possible corroboration of  
the equivalence between the CDT and RG approaches \cite{spectral-RG}.\footnote{
Inspired by the seemingly universal value of the UV spectral dimension, 
more general arguments about the underlying UV nature of space-time have been put forward
\cite{carlip}.}  

In view of the considerations outlined above, we have returned 
to a closer analysis of the basic CDT phase diagram. 
In what follows, we will report on some striking similarities between the
phase diagram of causal dynamically triangulated gravity and 
the Lifshifz phase diagram promoted in Ho\v rava-Lifshitz gravity. They become
apparent when one identifies ``average geometry", 
presumably related to the conformal mode of the geometry in some way,  
with the order parameter $\phi$ of 
an effective Lifshitz theory. We find that the phase structure 
allows potentially for both an anisotropic and an isotropic UV fixed point, opening
the exciting prospect that CDT can serve as a nonperturbative
lattice foundation for both the renormalization group approach and 
Ho\v rava-Lifshitz gravity, in the same way
as theories on fixed lattices provide us with nonperturbative definitions of 
quantum field theories in the formulation of K. Wilson.

\section{Causal Dynamical Triangulation}

We will merely sketch the setup used in CDT, and refer to \cite{d4,blp,bigs4} 
for more complete descriptions and to \cite{al} for the rationale behind the formulation.
We attempt to define the path integral of quantum 
gravity by summing over a class of piecewise linear spacetime geometries,
much in the same way as one can define the path integral in
ordinary quantum mechanics by dividing the time into intervals
of length $a$, considering paths which are linear between $t_n = na$
and $t_{n+1}= (n+1)a$, and then taking the limit of vanishing ``lattice spacing'', $a \to 0$.

Let us introduce a time slicing labeled by discrete lattice
times $t_n$. The spatial hypersurface labeled by $t_n$ has
the topology of $S^3$ and is a piecewise flat triangulation, obtained by gluing
together identical, equilateral tetrahedra with link length $a_s$, to be identified
with the short-distance lattice cut-off. 
We now connect the three-dimensional triangulation of $S^3$ at $t_n$
with that at time $t_{n+1}$ by means of four types 
of four-simplices: four-simplices of type (4,1), which share four vertices (in fact, an
entire tetrahedron) with the spatial hypersurface at $t_n$ and one vertex with the 
hypersurface at $t_{n+1}$; four-simplices of type (1,4), where the roles of 
$t_n$ and $t_{n+1}$ are 
interchanged; four-simplices of type (3,2), which share three vertices (in fact, an entire 
triangle) with the 
hypersurface at $t_n$, and two vertices with the hypersurface at $t_{n+1}$
(belonging to the same spatial link); lastly,
four-simplices of type (2,3), defined analogously but with $t_n$ and $t_{n+1}$ interchanged.

These four-simplices have a number of links (and corresponding triangles 
and tetrahedra) connecting vertices in hypersurfaces $t_n$ and $t_{n+1}$.
We take all of these links to be time-like with (squared) length $a^2_t = \a a^2_s$. 
The 
four-simplices are glued together such that the ``slab'' between hypersurfaces 
labeled by $t_n$ and $t_{n+1}$ 
has the topology $S^3\times [0,1]$. We {\it say} that 
the hypersurfaces are separated by a proper distance $\sqrt{\alpha}a_t$, but this 
is not strictly speaking true if one takes the piecewise flat geometries (despite their curvature
singularities) seriously as classical spacetimes. However, what is true is  
that all links connecting neighbouring hypersurfaces have proper length 
$\sqrt{\alpha}a_t$.

In the path integral we sum over all geometrically distinct piecewise linear geometries 
of this type, and with a fixed number of time steps.
As an action we use the Einstein-Hilbert action, which has a natural 
realization on piecewise linear geometries, first introduced by Regge.
The geometries allow a rotation to Euclidean geometries simply
by rotating $\a \to -\a$ in the lower-half complex plane. The action
changes accordingly and becomes the Euclidean Einstein-Hilbert Regge 
action of the thus ``Wick-rotated" piecewise flat Euclidean spacetime. 
Its functional form becomes extremely simple because we use only two 
different kinds of building blocks, which contribute in discrete units to
the four-volume and the scalar curvature.
In this way the Euclidean action becomes a function of ``counting building blocks",
namely,
\bea
S_E&=& \frac{1}{G} \int \sqrt{g} (-R+2\La) \nonumber \\
 &\to & -(\kappa_0+6\Delta) N_0+\kappa_4 (N_{4}^{(4,1)}+N_{4}^{(3,2)})+
\Delta (2 N_{4}^{(4,1)}+N_{4}^{(3,2)}),
\label{actshort}
\eea 
where $N_0$ is the number of vertices, $N_4^{(4,1)}$ 
the number of four-simplices of type (4,1) or (1,4),
and $N_{4}^{(3,2)}$ the number of four-simplices of type (3,2) or (2,3) in the
given triangulated spacetime history. For later use we 
denote the total number $N_4^{(4,1)}+N_4^{(3,2)}$ of four-simplices by $N_4$.
Furthermore, the parameter $\k_0$ in \rf{actshort} is proportional to the inverse bare
gravitational coupling constant, while $\k_4$ is related to the bare
cosmological coupling constant. Finally, $\Del$ is an asymmetry parameter
which in a convenient way encodes the dependence of the action on the relative
time-space scaling $\alpha$ introduced above, and is handy when studying 
the relation to Ho\v rava-Lifshitz gravity.  
Vanishing $\Del=0$ implies $\a =1$, and increasing $\Del$ away from zero 
corresponds to decreasing $\a$, i.e.\ the time-like links shrink in length when $\Del$ 
is increased. 

The rotation to Euclidean space is necessary in order to use Monte Carlo simulations
as a tool to explore the theory nonperturbatively.
For simulation-technical reasons it is preferable to keep the total 
number $N_4$ of four-simplices fixed during a Monte Carlo simulation, which
implies that $\kappa_4$ effectively does not appear as a coupling constant. 
Instead we can perform
simulations for different four-volumes if needed. To summarize, we are dealing with a
statistical system of fluctuating four-geometry, whose phase diagram as function of
the two bare 
coupling constants $\k_0$ and $\Del$ we are going to explore next.

\section{The CDT phase diagram}

\begin{figure}[t]
\center
\scalebox{0.57}{\rotatebox{-90}{\includegraphics{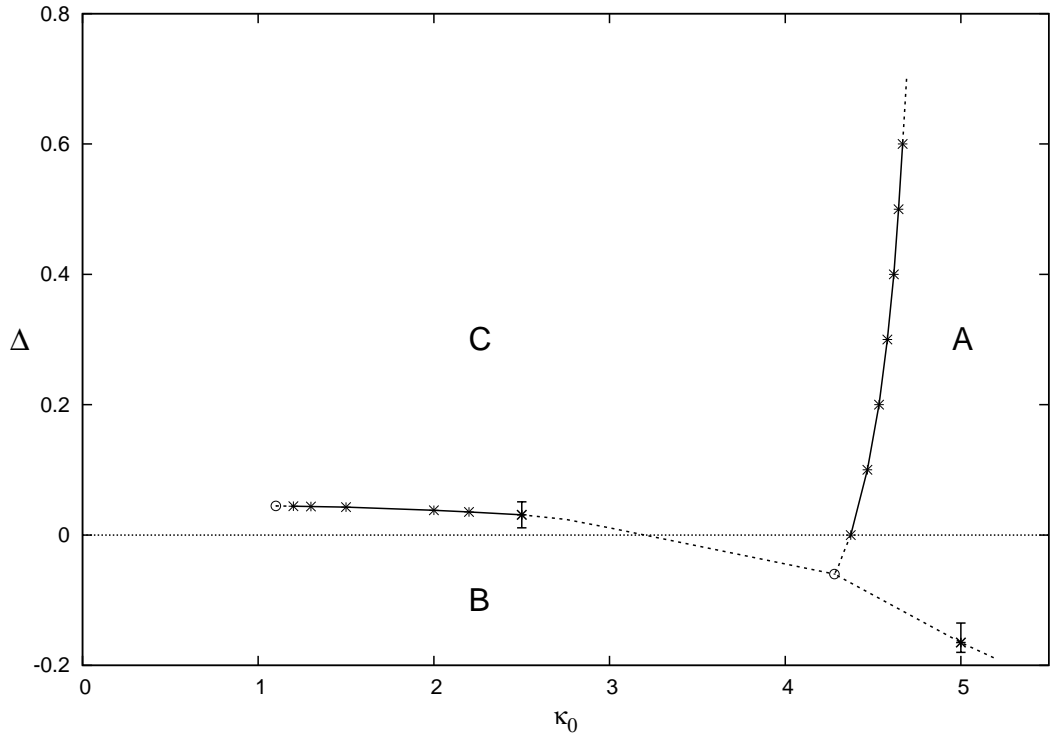}}}
\caption{The phase diagram of four-dimensional quantum gravity, defined
in terms of causal dynamical triangulations, parametrized by the inverse
bare gravitational coupling $\kappa_0$ and the asymmetry $\Delta$.} 
\label{fig1}
\end{figure}

The CDT phase diagram was described qualitatively as part of
the first comprehensive study of four-dimensional CDT quantum gravity \cite{blp}.
For the first time, we are presenting here the real phase diagram (Fig.\ \ref{fig1}), 
based on computer simulations with $N_4=80.000$.
Because there are residual finite-size effects for universes of this size, 
one can still expect minor changes in the location of the transition lines 
as $N_4 \to \infty$. The dotted lines in Fig.\ \ref{fig1} 
represent mere extrapolations,
and lie in a region of phase space which is difficult to access due
to inefficiencies of our computer algorithms. 

There are three phases, labeled A, B and C in \cite{blp}. 
In phase C, which had our main interest in \cite{bigs4,agjl}, we 
observed a genuine four-dimensional universe in the sense that 
as a function of the continuum four-volume $V_4$ (linearly related to
the number of four-simplices),
the time extent scaled as $V^{1/4}_4$ and the spatial volume 
as $V_4^{3/4}$. Moving into phase A, these 
scaling relations break down. Instead,
we observe a number of small universes arranged along the time 
direction like ``pearls on a string", if somewhat uneven in size.
They can split or merge along the time direction
as a function of the Monte Carlo time used in the simulations.
These universes are connected by thin ``necks'', i.e. slices 
of constant integer time $t_n$, 
where the spatial $S^3$-universes are at or close to the smallest
three-volume permitted (consisting of five tetrahedra glued together),
to prevent ``time" from becoming disconnected. 

By contrast, phase B is characterized by the ``vanishing" of the time direction,
in the sense that 
only one spatial hypersurface has a three-volume appreciably larger than 
the minimal cut-off size of five just mentioned. 
One might be tempted to conclude that the resulting universe is 
three-dimensional, just lacking the time direction of the extended
universe found in phase C. However, the situation is
more involved; although we have a large three-volume collected at
a single spatial hypersurface, the corresponding spatial universe has almost 
no extension. This follows from the fact (ascertained through measurement)
that it is possible to get in just a few steps from one tetrahedron to any other
by moving along the centres of neighbouring 
tetrahedra or, alternatively,
from one vertex to any other along a chain of links. 
The Hausdorff dimension is therefore quite high, and possibly infinite. 
Let us assume for the moment that it is indeed infinite; then the universe 
in phase B has neither time nor spatial extension, and there is
no geometry in any classical sense.  

We can now give the following qualitative characterization of the three
phases in terms of what we will provisionally call ``average geometry". 
The universe of phase C exhibits a classical four-dimensional background 
geometry on large scales, such that $\la {\it geometry}\ra \neq 0$. 
One may even argue that  $\la {\it geometry}\ra = const.$  
in view of the fact that according to the mini-superspace analysis of 
\cite{agjl,bigs4,semi} and allowing for a finite rescaling of the
renormalized proper time, the universe can be identified with the round 
$S^4$, a maximally symmetric de Sitter space of constant scalar curvature. 
By contrast, in phase B the universe presumably has no extension or
trace of classicality, corresponding to $\la {\it geometry}\ra = 0$.
Lastly, in phase A, the geometry of the universe appears to be
``oscillating'' in the time direction.
The three phases are separated by three phase transition lines which meet
in a triple point as illustrated in Fig.\ \ref{fig1}.

We have chosen this particular qualitative description to match precisely 
that of a Lifshitz phase diagram \cite{lifshitz}. In an effective Lifshitz theory, 
the Landau free energy density $F(x)$ 
as function of an order parameter $\phi(x)$
takes the form\footnote{see, for example, \cite{gold} for an introduction to
the content and scope of ``Landau theory"} 
\beq\label{2.2}
F(x) = a_2 \phi(x)^2 + a_4 \phi(x)^4 +a_6\phi(x)^6 + \ldots 
+c_2(\prt_\a \phi)^2 +d_2 (\prt_\b \phi)^2
+ e_2 (\prt_\b^2 \phi)^2 +\ldots ,
\eeq
where for a $d$-dimensional system $\a =m+1,\ldots,d$, $\b=1,\ldots,m$.
Distinguishing between ``$\alpha$"- and ``$\beta$"-directions allows
one to take anisotropic behaviour into account.
For a usual system, $m=0$ and a phase transition can occur when 
$a_2$ passes through zero (say, as a function of temperature).
For $a_2> 0$ we have $\phi=0$, while for $a_2 <0$ we have $|\phi| >0$ 
(always assuming $a_4 >0$).  However, one also has a transition
when anisotropy is present ($m>0$) and 
$d_2$ passes through zero. For negative $d_2$
one can then have an oscillating behaviour of $\phi$ in the $m$ 
``$\beta$"-directions.
Depending on the sign of $a_2$, the transition to this 
so-called modulated or helical phase can occur either from the 
phase where $\phi=0$, or from the phase where $|\phi| >0$. 
We conclude that the phases C, B, and A 
of CDT quantum gravity depicted in
Fig.\ \ref{fig1} can be put into a one-to-one correspondence 
with the ferromagnetic, paramagnetic and helical phases
of the Lifshitz phase diagram\footnote{For 
definiteness, we are using here a ``magnetic'' language for the Lifshitz 
diagram. However, the Lifshitz diagram can 
also describe a variety of other systems, for instance, liquid crystals.}. 
The triple point
where the three phases meet is the so-called Lifshitz point.

The critical dimension beyond which the mean-field 
Lifshitz theory alluded to above is believed to be valid is 
$d_c = 4+m/2$. In lower dimensions, the fluctuations play an
important role and so does the number of components of the field $\phi$.
This does not necessarily affect the general structure of the phase
diagram, but can alter the order of the transitions.
Without entering into the details of the rather complex general situation,
let us just mention that for $m=1$ fluctuations will often turn the
transition along the A-C phase boundary into a first-order transition.
Likewise, most often the transition between phases B and C is 
of second order. 

We have tried to determine the order of the transitions in the CDT 
phase diagram. Based on our numerical investigation so far, the 
A-C transition appears to be 
a first-order transition, while the B-C transition may be either of 
first or second order. This points to significant 
similarities with the Lifshitz results mentioned above, 
although we would like
to stress that at this stage this is only at the level of analogy. 
We have not yet derived a gravitational analogue of the Landau free
energy \rf{2.2} governing the effective Lifshitz model. Also, a difference
which may turn out to be important is that in our case the B-C phase 
transition line seems to end, and the endpoint may play a special role, 
as we will discuss below.

\begin{figure}[t]
\center
\scalebox{0.57}{\rotatebox{-90}{\includegraphics{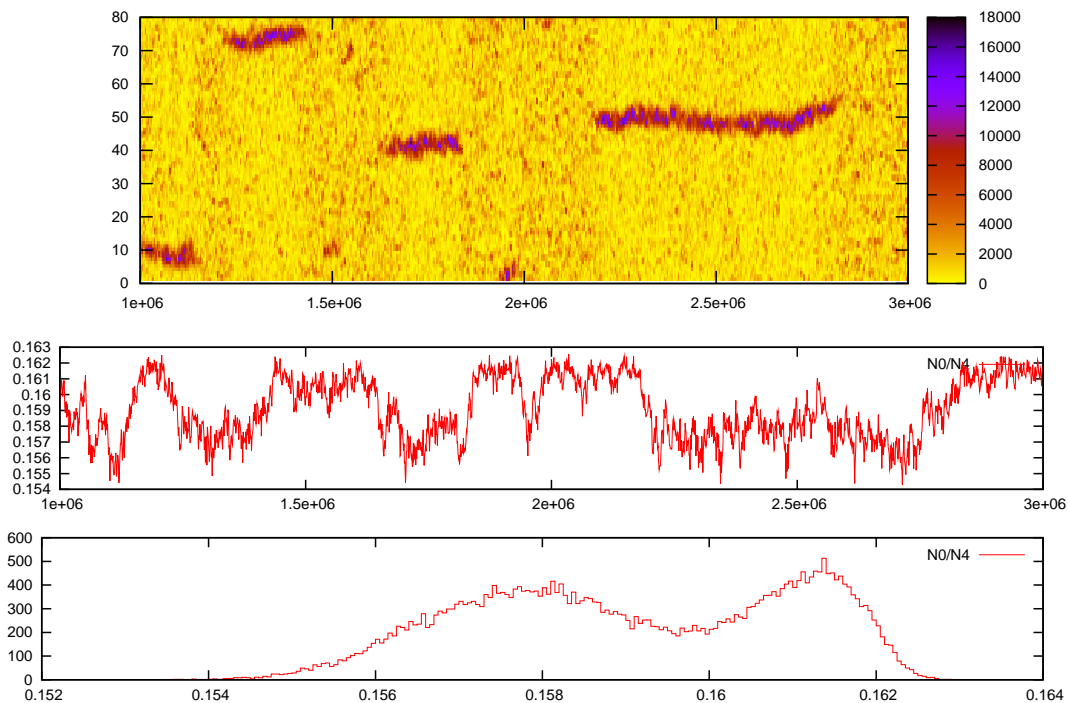}}}
\caption{Evidence for a first-order signal at the A-C phase transition.} 
\label{fig2}
\end{figure}

The two graphs at the bottom of Fig.\ \ref{fig2} illustrate the behaviour
of $N_0/N_4$ at the A-C phase transition line. 
Since we can approach this line by changing the coupling constant $\k_0$
while keeping $\Del$ fixed, the quantity conjugate to $\kappa_0$ 
($N_4$ is fixed), namely, the ratio $N_0/N_4$, is a natural
candidate for an order parameter. The graph at the centre of Fig.\ \ref{fig2} 
shows $N_0/N_4$ as a function of Monte Carlo time.
One sees clearly that it jumps between two values, corresponding to
the distinct nature of geometry in phases A and C. We have checked that 
the geometry indeed ``jumps" in the sense that no 
smoothly interpolating typical configurations have been found. Lastly,
we have also established that the jump becomes more and more 
pronounced as the four-volume $N_4$ of the universe increases, further
underlining the archetypical first-order behaviour at this transition line. 

The top graph in Fig.\ \ref{fig2} shows the location of the universe
along the vertical ``proper-time
axis" ($t_n\in[ 0, 80]$, and to be periodically identified) as a function of
Monte Carlo time, plotted along the horizontal axis. 
The value of the spatial three-volume $V_3(t_n)$ in the slice labelled by 
$t_n$ is colour-coded; the darker, the bigger the volume at time $t_n$.
We can distinguish two types of behaviour as a function of Monte Carlo time,
(i) presence of an extended universe centred at and fluctuating weakly around 
some location on the proper-time axis; (ii) absence of such a universe with
a well-defined ``centre-of-volume". The former is associated with the presence
of a distinct dark band in the figure, which disappears abruptly as a function
of Monte Carlo time, only to reappear at some different location $t_n$ later
on in the simulation. Comparing with the middle graph, it is clear that 
these abrupt changes in geometry correlate perfectly with 
the changes of the order parameter $N_0/N_4$. 
When $N_0/N_4$ is small, we witness the extended de Sitter
universe of phase C,
whose ``equator" coincides with the dark blue/red line of the colour plot.
Conversely, at the larger values of $N_0/N_4$ characteristic of phase A 
this structure disappears, to be replaced by an array of universes
too small to be individually identifiable on the plot. When jumping back
to phase C the centre-of-volume of the single, extended universe 
reappears at a different location in time. 
Finally, the bottom graph in Fig.\ \ref{fig2} illustrates the double-peak 
structure of the distribution of the values taken by the order 
parameter $N_0/N_4$.

\begin{figure}[t]
\center
\scalebox{0.57}{\rotatebox{-90}{\includegraphics{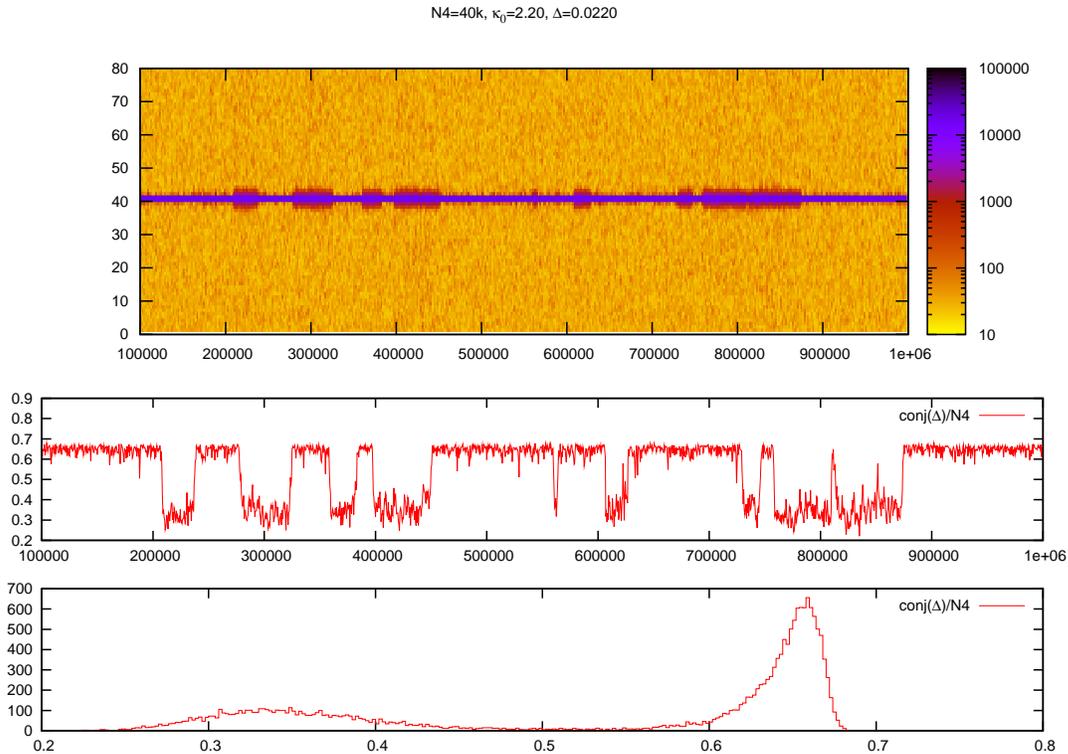}}}
\caption{Evidence for a first-order signal at the B-C phase transition -- or is it?} 
\label{fig3}
\end{figure}

Our measurements to determine the character of the B-C transition are
depicted in an analogous manner in Fig.\ \ref{fig3}. 
Since we are varying $\Delta$ to reach this transition from inside phase C, 
we have chosen the variable conjugate to $\Delta$ in the action \rf{actshort} (up
to a constant normalization $N_4$),
$(-6N_0 +N_4^{(4,1)})/N_4$, as our order parameter.
Looking at the graph at the centre, we see that this parameter exhibits the same
jumping behaviour as a function of Monte Carlo time characteristic of a 
first-order transition. Small values of the parameter indicate the system is
in phase C, while large values correspond to phase B.
The time extent of the universe diminishes
as one approaches the phase transition line from phase C, 
but it does not go to zero. It jumps to zero only when we cross the line.
Some indication of this behaviour is given by the colour-coded three-volume 
profile $V_3(t)$ as function of the Monte Carlo time in the top graph of 
Fig.\ \ref{fig3}. In phase B, the entire ``universe" is concentrated in a single
slice, while in phase C it has a non-trivial time extension. 
The bottom graph in Fig.\ \ref{fig3} again 
shows the double-peak structure of the order parameter. 

Looking at Fig.\ \ref{fig3} and comparing it with the previous Fig.\ \ref{fig2},
the evidence for a first-order transition at the
B-C phase boundary seems even more clear cut than in the case of the
A-C transition. However, there is one set of measurements which potentially
calls this result into question. By studying the strength of this signal systematically 
as a function of the total four-volume $N_4$, we found that it becomes weaker with
increasing $N_4$; the sharp jumps seen in Fig.\ \ref{fig3} become blurred 
and the two peaks of the order parameter start to merge.  
We should therefore keep an open mind to the possibility that the
observed behaviour is an artifact of using systems which are too small to see
their true nature. 
Unfortunately it is presently not possible for us to use much larger
systems, because the local Monte Carlo algorithm becomes quite 
inefficient as one approaches the transition, and the autocorrelation time 
grows rapidly with the four-volume.  
  
We have not studied the A-B transition in any detail since it seems
not interesting from a gravitational point of view, where we want to start
out in a phase with an extended, quasi-classical four-dimensional universe,
i.e.\ in phase C.

Finally, the order parameters used above, given in terms of the 
conjugates to the coupling constants varied in the action \rf{actshort}
cannot be identified directly with the ``average geometry"
alluded to in the introduction, although they must clearly capture some
aspects of it. Some further speculations on the nature of the
order parameter can be found in Sec.\ \ref{disc} below.

\section{Relation to Ho\v rava-Lifshitz and RG gravity}

When we approach the B-C phase transition line from phase $C$ by varying 
the coupling constant $\Del$, the time extent of the universe becomes smaller, 
when measured in units of discrete time steps (recall that we are keeping
the four-volume constant). However, as argued in \cite{round}, the real time 
extent does not change all that much, since there is a compensating effect 
of ``stretching" of individual four-simplices in the time direction due to an
increase in $\alpha$, the relative scaling between space- and time-like
lattice spacings defined by $a_t^2 = \a a_s^2$ \cite{d4}, and set by hand.
(As mentioned above, for given values of $\kappa_0$ and $\kappa_4$ 
one can calculate $\Del$ as a function of $\a$.)
In addition, one needs to keep in mind that 
the precise effective action underlying the form of the phase diagram
in Fig.\ \ref{fig1} is truly nonperturbative, since it results from the interplay of
the bare action \rf{actshort} and the measure (i.e.\ the entropy of 
the geometries as a function of $N_0$, $N_4^{(3,2)}$, $N_4^{(4,1)}$), and is 
therefore difficult to calculate. As a consequence, it may happen 
that the effective action changes {\it form} 
when we change $\k_0$ or even $\Del$. Since the CDT geometries, unlike
their Euclidean counterparts, explicitly
break the isotropy between space and time, a natural 
deformation could be ``Ho\v rava-Lifshitz-like''. In this case
a potential effective Euclidean continuum action, including the measure, 
and expressed
in terms of standard metric variables could be of the form
\beql{horava}
S_H = \frac{1}{G} \int \d^3x\ \d t \; N \sqrt{g}
\Big((K_{ij}K^{ij}-\lam K^2) + (-\g R^{(3)} +2\La + V(g_{ij})\Big),
\eeq
where $K_{ij}$ denotes the extrinsic curvature, $g_{ij}$ the three-metric 
on the spatial slices, $R^{(3)}$ the corresponding 3d scalar 
curvature, $N$ the lapse function, and finally $V$ a ``potential'' which in Ho\v rava's 
formulation would contain higher orders of spatial derivatives, potentially 
rendering $S_H$ renormalizable. The kinetic term depending on the extrinsic
curvature is the most general such term which is at most second order in 
time derivatives and consistent with spatial diffeomorphism invariance.
The parameter $\lambda$ appears in the (generalized) DeWitt metric, which 
defines an ultralocal metric on the classical space of all 
three-metrics.\footnote{The value of $\lambda$ governs the signature of
the generalized DeWitt metric 
$$
G_\lambda^{ijkl}=\frac{1}{2}\sqrt{\det g} (g^{ik} g^{jl}+g^{il}g^{jk}-
2\lambda g^{ij} g^{kl}),
$$ which is positive definite for $\lambda <1/3$, indefinite
for $\lambda =1/3$ and negative definite for $\lambda >1/3$.}
The parameter $\gamma$ can be related to a relative scaling between 
time and spatial directions.
When $\lam =\g =1$ and $V=0$ we recover the standard (Euclidean)
Einstein-Hilbert action. 

Making a simple mini-superspace ansatz with compact spherical slices, 
which assumes homogeneity and isotropy of the spatial three-metric $g_{ij}$,
and fixing the lapse to $N=1$, the Euclidean action (\ref{horava}) becomes 
a function of the scale factor $a(t)$ 
(see also \cite{elias,brandenberger,calcagni}), that is,
\beql{mini}
S_{mini} = \frac{2 \pi^2}{G} \int \d t \; a(t)^3 \Big( 3(1-3\lambda)\ 
\frac{\dot{a}^2}{a^2} -\gamma\ \frac{6}{a} +2 \Lambda+ \tV(a)\Big).
\eeq 
The first three terms in the parentheses define the 
IR limit, while the potential term $\tV(a)$ contains  
inverse powers of the scale factor $a$ coming from possible
higher-order spatial derivative terms.  

The results reported in \cite{bigs4,agjl,semi} 
are compatible with the functional form of the mini-superspace action
\rf{mini}, but we were not able to determine $\tV(a)$, which could 
be important for small values of the scale factor.
As outlined in \cite{bigs4}, because of the nature of the dependence of
the {\it renormalized} coupling constants on the bare parameters $\kappa_0$,
$\Delta$, resolving shorter distances seems to necessitate a closer
approach to the phase transition lines, with   
UV behaviour found along those lines. The A-C line
is first order and thus not of interest. The order of the A-B line
has not been determined, but since it cannot be approached from 
inside phase C which has good IR properties, it is currently not of
interest either. This leaves the B-C transition line, whose character has
not yet been established. {\it If}\, it is a first-order line, we are 
left with two potentially interesting points: the endpoint $P_0$ of the transition, 
where the phase transition is often of higher order than along the line itself,
and the Lifshitz triple point $P_t$, where the transition also might be 
of higher order. On the other hand, if the line is second order, we can 
approach it anywhere.

One defining aspect of Ho\v rava-Lifshitz gravity is the assumption that 
the scaling dimensions of space and time differ in the 
ultraviolet regime. This difference is used to construct a theory 
containing higher spatial derivatives in such a way that it
is renormalizable. How would one observe such a difference 
in the present lattice approach? Consider a universe of time 
extent $T$, spatial extension $L$ and total four-volume $V_4(T,L)$.
By measuring $T$ and $L$ we can establish the mutual relations
\beql{5.1}
T \propto V_4^{1/d_t},~~~~L\propto \Big(V_4^{1-1/d_t}\Big)^{1/d_s}\propto 
T^{(d_t-1)/d_s}.
\eeq
Well inside phase $C$ we measured $d_t=4$ and $d_s=3$, in agreement
with what is expected for an ordinary four-dimensional space-time.
If the dimension [T] of time was $z$ times the dimension [L] of length
we would have
\beql{5.2}
z= \frac{d_s}{d_t-1}.
\eeq
We have seen that well inside phase B both $d_s$ and $d_t$ must
be large, if not infinite. If the B-C phase transition is second order, it may 
happen that $z$ goes to a value different from 1 when we approach
the transition line. To investigate this possibility, we have tried to 
determine $z$ as a function 
of the parameter $\Del$ as $\Delta\rightarrow 0$.
For $\Del > 0.3$ one obtains convincingly $d_t \approx4$ and 
$d_s \approx 3$ and thus $z\approx 1$, but for smaller $\Del$ 
the quality of our results does not allow for any definite statements. 
Autocorrelation times seem to become very long and there may be large 
finite-volume effects, which obscure the measurements and which are 
precisely based on finite-size scaling. Hopefully the 
latter are more a function of our algorithms than
indicating a need to go to much larger four-volumes.

\section{Discussion}\label{disc}

We have shown that the CDT phase diagram bears a 
striking resemblance to a Lifshitz phase diagram
if we identify ``average geometry" with the Lifshitz field $\phi$
in the heuristic sense discussed above.
In our earlier papers, we tentatively interpreted phase A as  
being dominated by the wrong-sign kinetic term of the conformal factor 
of the continuum Euclidean Einstein-Hilbert action.
It appears to be a Lorentzian remnant of a degeneracy found
in the old Euclidean quantum gravity model based on (Euclidean) 
dynamical triangulations, which for large values of $\k_0$ exhibits
a branched-polymer phase, likewise interpreted as caused by the 
dominance of the conformal factor (see, for example, \cite{ambjur}). 

Eq.\ \rf{2.2} strongly suggests an identification of the 
Lifshitz field $\phi$ with the conformal factor or some function thereof,
such that the transition from phase C to phase A in the Lifshitz diagram 
is associated with a sign swap of the corresponding kinetic term
from negative to positive. An effective action for the
conformal mode coming out of a nonperturbative gravitational path integral    
would consist of (i) a contribution from the bare action (where the kinetic
conformal term has the ``wrong", negative sign), (ii) a contribution from the 
measure, and (iii) contributions from integrating out other field components
and, where applicable, other matter fields. It has been argued previously
that the Faddeev-Popov determinants contributing to the
gravitational path integral after gauge-fixing contribute effectively to the conformal
kinetic term with the opposite, positive sign \cite{mm,dl}. 
For example, when working in proper-time
gauge, to imitate the time-slicing of CDT, Euclidean metrics can be
decomposed according to\footnote{The conformal decomposition of the
spatial three-metric is essentially unique if one requires $g_{ij}$ to have constant
scalar curvature.}
\begin{equation}
ds^2=d\tau^2+{\rm e}^{2\phi(\tau,x)} g_{ij}(\tau,x)dx^i dx^j,
\end{equation}
giving rise to a term $-1/G^{(b)} {\rm e}^{3\phi}\sqrt{\det g} (\partial_\tau\phi)^2$ 
in the bare gravity Lagrangian density, where $G^{(b)}$ is the bare Newton's constant.
According to \cite{dl}, one expects that the leading contribution from the associated
Faddeev-Popov determinant has the same functional form, but with a
plus instead of a minus sign, and with a different dependence on $G^{(b)}$. 
The presence of contributions of opposite sign 
to the effective action for the conformal mode
$\phi(\tau,x)$ can therefore
lead to two different behaviours, depending on the value of
$G^{(b)}$. Identifying the $\kappa_0$ of our lattice formulation -- an a priori freely
specifiable parameter -- with the
inverse of $G^{(b)}$, this mechanism can account exactly for the observed
behaviour with regard to the transition between phases A and C.\footnote{Related
mechanisms have earlier been considered in the 
context of purely Euclidean dynamical triangulations by J.\ Smit \cite{js}.}

However, we have at this stage not constructed a gravity analogue of the
Landau free energy density (\ref{2.2}) incorporating all
observed features of the CDT phase diagram.
Also, as already mentioned above, due to finite-size effects and/or 
the inefficiency of our computer algorithms in this region
we have not yet been able to establish the order of the B-C phase transition.
In a Lifshitz diagram it is often of second order. 
If this scenario was realized in CDT quantum gravity too,
any point on the line could could potentially be associated with a 
continuum UV limit, thus implying the need for fine-tuning 
at least one other coupling constant (apart from the cosmological constant).

Engaging in a bit of speculation, an interesting possibility would be if 
the triple Lifshitz point corresponded to an asymmetric scaling 
between space and time, like in the Ho\v rava model, while 
the endpoint of the B-C line represented an isotropic point 
associated with the RG asymptotic safety picture.
Deciding which scenario is actually realized will
require substantially longer simulations or improved
updating algorithms, something we are working on presently. --
The analysis of the phase structure in the framework of causal
dynamical triangulations we have performed here may turn out to
provide a universal template for understanding nonperturbative
theories of higher-dimensional, dynamical geometry,
including ``true" quantum gravity.

\vspace{.5cm}

\noindent{\bf Acknowledgment.} 
JJ acknowledges partial support by the Polish Science Foundation's
International Ph.D. Projects Programme, co-financed
by the European Regional Development Fund under agreement no.
MPD/2009/6.
RL acknowledges support by the Netherlands
Organisation for Scientific Research (NWO) under their VICI
program. In addition, the contributions by SJ and RL to this work are 
part of the research programme of the 
Foundation for Fundamental Research on Matter (FOM), financially 
supported by NWO.

\end{document}